# Galaxies with the DENIS 2 Micron Survey: A Preliminary Report

Gary A. Mamon[1,2]

[1] Institut d'Astrophysique, 98 bis Blvd Arago, F–75014 Paris, France
[2] DAEC, Observatoire de Paris, F–92195 Meudon, France

**Abstract.** The DENIS *IJK* survey of the entire southern sky is beginning observations this year. It will return one million images, representing 4 Terabytes, among which nearly $10^8$ stars and up to one million galaxies. The main extragalactic applications of such a multi-band, digital, near-IR 2D survey are listed, and the importance and feasibility of a spectroscopic followup are discussed. First estimates of the capabilities for galaxy extraction are given, and illustrated with preliminary images of three very bright galaxies.

## 1 Introduction

The DENIS (DEep Near-Infrared Survey of the Southern Sky) survey is a European effort to map the entire southern sky ($-88° \leq \delta \leq +2°$) in the $I$ (0.8 µm), $J$ (1.25 µm), and $K_s$ (2.15 µm) bands (see Epchtein et al. 1994; Ruphy, in these proceedings). The survey uses the ESO 1m telescope, fulltime, and is expected to take 3 or 4 years. DENIS should produce one million images in each of the 3 bands (observed concurrently), yielding nearly one hundred million stars, and up to one million galaxies (see Sect. 3 and Table 1 below). The 4 Terabytes of images will be analyzed partly on site in Chile, partly in Paris and partly in Leiden. An analogous effort is being taken by the American 2MASS project (see Lonsdale, in these proceedings), which will map *both* hemispheres, with dedicated telescopes, in the $J$, $H$, and $K_s$ bands. DENIS is scheduled to commence in the fall of 1995 (the $I$ camera has only been on-line since July 1995), while 2MASS is expected to begin the Northern survey in early 1997. The reader is referred to Ruphy (in these proceedings) for additional details and results from preliminary star counts.

## 2   Extragalactic Applications

As emphasized throughout this meeting, the Near-Infrared (NIR) has two important advantages for extragalactic and cosmological science:

1) The NIR bands lie represent the optimum region of the electromagnetic spectrum between *low dust extinction* and *low dust emissivity*.
2) The NIR light is expected to be better correlated with the *mass of the stellar component* of galaxies than the optical light (enhanced by recent star formation) or Mid- to Far-Infrared light (tracing warm dust, associated with recent star formation).

The first advantage allows one to probe galaxies behind the plane of the Milky Way (e.g. Mamon 1994), which is all the more important that the most important concentration of mass in the local Universe, the *Great Attractor*, seems to be centered at $b = 0°$ (Kolatt, Dekel and Lahav 1995). In comparison, optical surveys (e.g. CfA2, SSRS2) are typically limited to $|b| > 20°$, while IRAS is limited at $|b| \leq 12°$, mainly by stars and cirrus (Meurs and Harmon 1988). Moreover, the near transparency of dust to NIR light allows one to have a full view of external galaxies.

It seems reasonable to extrapolate the second advantage to the fact that NIR light should trace best the underlying total mass distribution of the Universe, hence the importance of NIR surveys for cosmological analyses of the large-scale structure of the Universe.

With these advantages in mind, the applications of such 2D surveys as DENIS and 2MASS are

1) A large (few thousand) set of bright galaxies from which one will obtain statistics of their morphological features, such as profiles in surface brightness, color, axis-ratio, and position angle. This will be especially true with the *I*-band images, for which the spatial resolution is superior, and hence a large enough database of galaxies with enough details.
2) The 2D distribution of galaxies in each of the three bands. In particular, Establishment of catalogs of binaries, loose groups, compact groups, and clusters
   a) Quantification of large-scale structure (2-point correlation functions, counts in cells ...)
   b) Mapping of large-scale structure through the Galactic Plane, in particular to observe the large but only slightly overdense Great Attractor.
3) The relation between color and environment, i.e. *color-segregation*, which itself is related to morphology and bulge/disk ratio.
4) The verification of galaxy counts at the bright end (where current surveys usually suffer from low statistics, see Gardner, Cowie, and Wainscoat 1993).

In addition, 2MASS, with its full-sky coverage, should probe the cosmic dipole to estimate the acceleration of the Local Group.

A spectroscopic followup will be necessary to obtain

1) A 3D view of the local Universe
2) Internal kinematics of groups and clusters

Unfortunately, it is a costly proposition to measure tens to hundreds of thousands of galaxy redshifts. A comparison of the efficiency of different instruments available in the southern hemisphere (Mamon 1995) indicates that the 2dF 400-fiber spectrograph on the 4m AAT is by far the most efficient instrument, only requiring 200 ($B < 17$) to 400 ($B < 19$) clear nights to cover a hemisphere of galaxies (without regard of the extinction near the Galactic Plane). However, in view of the pressure on the 2dF and the shallowness of the DENIS limiting magnitudes (the targets are so sparse that the 2dF would only use a fraction of its fibers) it is not realistic to expect much observing time for a spectroscopic followup of DENIS. The alternative would be to use an upgrade of the FLAIR-II 100-fiber spectrograph mounted on the UKST Schmidt telescope (see Parker and Watson 1995), which, with additional fibers and automated fiber-positioning, could be faster than the 2dF for a shallow ($B_{\rm lim} = 17$) spectroscopic followup. It is subject to lower user-pressure than the 2dF and a 300-fiber version of FLAIR would use all of its fibers at $B = 17$, which is the rough equivalent blue limit for a $K$-catalog obtained with cooled optics. which thus matches well the (see Table 1 below).

## 3  DENIS Capabilities for Galaxy Extraction

The capabilities of DENIS for detecting galaxies with decent photometry and separating them from stars have been estimated first with simulated images, and will later be checked by comparing with sets of 6-times coadded images (see Mamon *et al.* 1996 for more details on the strategy for optimizing the algorithms and estimating the magnitude limits). We will effectively be limited by the accuracy of the photometry for spiral galaxies and by star/galaxy separation for ellipticals. The preliminary magnitude limits and the catalog sizes we expect are listed in Table 1 below, where completeness and reliability are required at the 95% level *for each galaxy type* not explicit in the last column. The $K$ limits are expected to improve with the projected cooling of the optics (only the detectors currently sit within dewars), which could lead to catalogs 3 times larger in $K$.

**Table 1.** DENIS galaxy catalogs with 95% completeness/reliability

| Magnitude limit | Size ($2\pi$ sr) | Incompleteness ($> 5\%$) |
|---|---|---|
| $K < 12.2$ | 30 000 | $|b| < 2°$, Face-on Sc→Ir |
| $J < 14.4$ | 110 000 | $|b| < 5°$ |
| $J < 15.6$ | 300 000 | $|b| < 5°$, Ellipticals |
| $I < 17.0$ | 1 100 000 | $|b| < 10°$ |

## 4   Bright Galaxies

In Figure 1 are shown the 3 brightest galaxies (NGC 1512: SBa, $B_T = 11.1$; NGC 1515: Sbc, $B_T = 12.0$; NGC 1527: S0, $B_T = 11.7$) of the RC3 catalog (de Vaucouleurs *et al.* 1991) that were in the regions available on computer-disk at the DENIS Paris Data Analysis Center in Sept. 1994. The figure illustrates galaxies in the DENIS $J$ and $K_s$ bands ($9 \times 1$ s exposures), and for comparison, in 70 min (NGC 1512 and NGC 1515) and 75 min (NGC 1527) exposures in blue ($b_J$) on the UKST Schmidt telescope (obtained from the Digitized Sky Survey).

In the DENIS $J$ image, the bar of NGC 1512 is clearly seen with hints of the spiral arms starting at the edge of the bar. In the corresponding $K_s$ image, both the bar and the beginning of the spiral pattern are only very marginally seen, whereas the nucleus is clearly visible. The disk of NGC 1512, seen nearly face-on, is invisible in both $J$ and $K_s$.

The $J$ image of the nearly edge-on spiral NGC 1515 traces the inner disk with hints of the inner spiral structure. In $K_s$, the galaxy is much smaller, with only the smallest hints of the inner spiral structure.

Finally, the nearly edge-on lenticular NGC 1527, being nearly featureless in the optical is of course even more so in the DENIS images, only the inner nucleus and the beginning of the disk are seen in $K_s$.

Altogether, these 3 images illustrate 1) the greater sensitivity of $J$ relative to $K_s$ in DENIS, which is mainly caused by the much higher atmospheric/instrumental background in the latter band; 2) very bright galaxies $B_T \leq 12$ can show bars and spiral arms in the DENIS $J$ and $K_s$ bands.

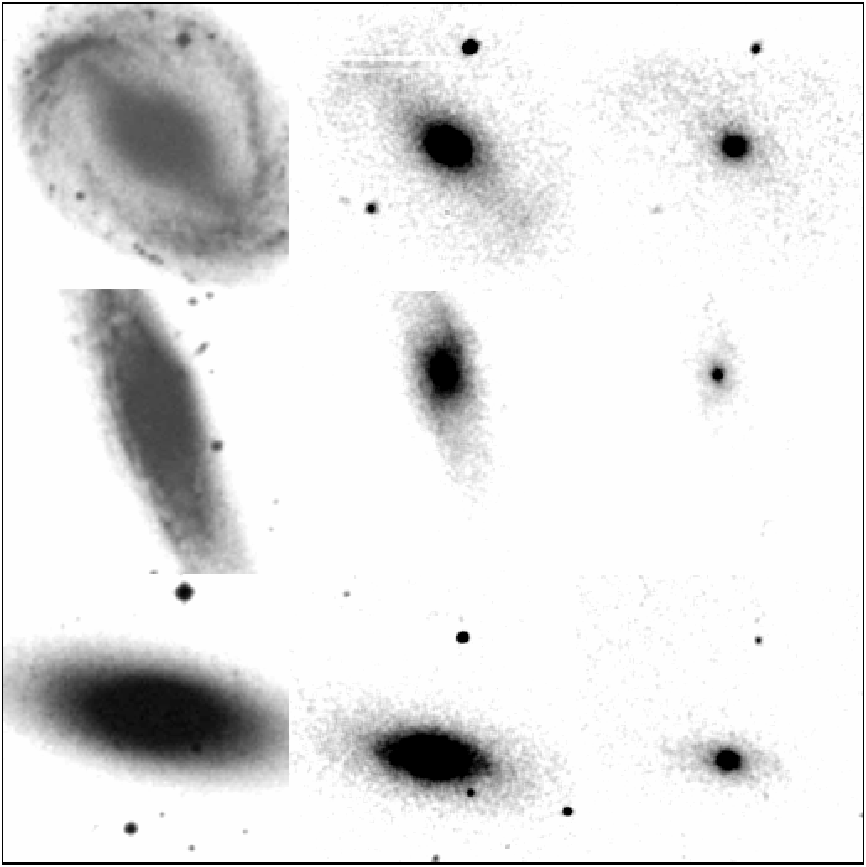

**Figure 1.** Three bright galaxies seen in $b_J$ (*left* from Digitized Sky Survey) and the DENIS $J$ (*middle*) and $K_s$ (*right*) bands; *top:* NGC 1512, *middle:* NGC 1515, and *bottom:* NGC 1527. Fields are $6'$ wide. NIR images are $3\times3$ boxcar smoothed. Greyscales (white to black) are $9 \to 60\sigma$ ($b_J$) and $2 \to 20\sigma$ ($J$ and $K_s$, where $\sigma$ is after smoothing). Because NGC 1515 and 1527 are situated near the edges of the DENIS images, there is a small vertical misalignment with the optical images.